# Extraction-based Deep Learning Reconstruction of Interior Tomography

Changyu Chen[a,b], Yuxiang Xing[a,b], Li Zhang[a,b], and Zhiqiang Chen[a,b]

[a]Department of Engineering Physics, Tsinghua University, Beijing, 100084, China;

[b]Key Laboratory of Particle & Radiation Imaging (Tsinghua University), Ministry of Education, Beijing, 100084, China

## ABSTRACT

Interior tomography is a typical strategy for radiation dose reduction in computed tomography, where only a certain region-of-interest (ROI) is scanned. However, given the truncated projection data, ROI reconstruction by conventional analytical algorithms may suffer from severe cupping artifacts. In this paper, we proposed a new extraction-based deep learning method for the reconstruction of interior tomography. Our approach works in dual domains where a sinogram-domain network (SDNet) estimates the contribution of the exterior region to the truncated projection and an image-domain network (IDNet) further mitigates artifacts. Unlike the previous extrapolation-based methods, SDNet is intended to obtain a complete ROI-only sinogram via extraction instead of a fully non-truncated sinogram for both the ROI and exterior regions. Our experiments validated the proposed method and the results indicate that the proposed method can disclose more reliable structures. It achieved better image quality with better generalization performance than extrapolation-based methods.

**Keywords:** Computed tomography (CT), deep learning, interior tomography, sinogram extraction

## 1. INTRODUCTION

X-ray computed tomography (CT) plays a key role in clinical diagnosis while the potential risk of radiation dose to patients is highly concerned. Hence, various strategies have been investigated for dose reduction. Interior tomography is commonly used when only a certain region of the patient is of interest. With X-ray flux collimated towards the region-of-interest (ROI), interior tomography can significantly reduce unnecessary dose exposure to the uninterested exterior region. However, projection data will be truncated due to the limited field-of-view (FOV) scanning. If the standard filtered back-projection (FBP) algorithm is employed for reconstruction, the truncated filtration will introduce severe cupping artifacts into the reconstructed ROIs which compromises important diagnostic information.

In last decades, a lot of work has gone into interior reconstruction. One straightforward method is sinogram extrapolation. These methods made certain assumptions for the exterior region and apply heuristic methods for extrapolation such as cosine fitting[1] and elliptical extrapolation[2]. Unfortunately, the reconstructed value of attenuation could be biased due to the inaccuracy of extrapolation. Other researchers explored exact interior reconstruction with various conditions of known subregions[3, 4] and developed reconstruction algorithms under the differentiated back-projection (DBP) framework[5]. With the advent of compressed sensing, the sparsity model of ROIs is integrated into reconstruction[6]. Given an ROI is piecewise constant or polynomial, it can be stably reconstructed via total variation (TV) or high-order TV minimization[6, 7], respectively. However, in practical situations, these methods may incur difficulties that known subregions are not always available and the sparsity model may be violated.

Recently, deep learning (DL) has yielded impressive performance in various ill-posed reconstruction problems. In the scope of interior tomography, some researchers propose to use the filtered backprojection (FBP)[8], DBP[9] reconstructions, or direct back-projected projection[10] as the input of U-Net for post-processing. Others focus on sinogram extrapolation [11] or combine it with an image-domain network to form a dual-domain optimization[12]. To achieve better robustness, an extrapolation-based plug-and-play method was also explored[13]. Although these pioneering methods have shown promising results for interior reconstruction, there are still some concerns to be addressed. On the one hand, the post-processing methods failed to integrate the physical model into the network which could be harmful to the robustness. On the other hand, the extrapolation-based methods learn to estimate the undetected projection to reflect the complete measure of both the ROI and the exterior region, which is very challenging especially when the ROI size is relatively small. Intuitively, the truncated projection data can be viewed as the combination of a complete measure of the ROI and a truncated measure of the exterior region. Hence, exterior reconstruction is highly underdetermined with a big exterior region. Besides, if the scanning is coupled with other nonideal factors such as high noise and metal artifacts, extrapolation-based methods may not be efficient enough to deal with such distortions inside the truncated projection.

In this paper, we proposed a new extraction-based dual-domain pathway for interior reconstruction. Specifically, a sinogram-domain network (SDNet) estimates the contribution of exterior region to the truncated projection, and an image-domain network (IDNet) further mitigates artifacts. Our experiments indicate that the proposed method can disclose more reliable structures and achieve better image quality with better capability of generalizing to a higher noise level than extrapolation-based methods.

## 2. METHODOLOGY

Without losing generality, we formulate interior tomography under fan-beam scanning (see Fig. 1(a)). Let $\mathbf{\mu}_{FULL}$ being the attenuation map of an object imaged with a ROI denoted by $\mathbf{\mu}_{ROI}$ and an exterior region denoted by $\mathbf{\mu}_{EX}$ (see Fig. 1(b)).

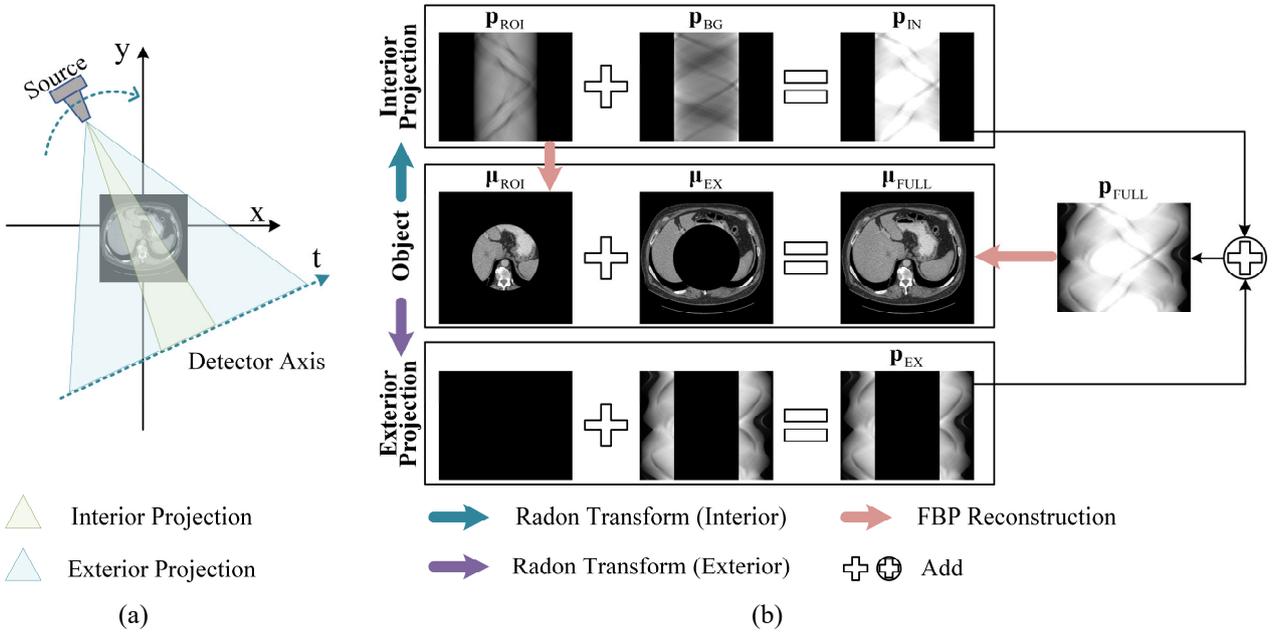

**Fig. 1** Illustration of Interior Tomography: the system configuration of interior projection (green triangle) and exterior projection (blue triangle) (a). Radon transform for interior projection (blue arrow) and exterior projection (purple arrow). FBP reconstruction is denoted by red arrows (b).

For interior tomography, an interior projection $\mathbf{p}_{IN}(\theta,t)$ is acquired with $\theta \in [0, 2\pi)$ being the angle of projection views and $t$ the coordinate of detector elements (see Eq. (1)). With X-ray flux collimated to fully cover $\mathbf{\mu}_{ROI}$, the field of view (FOV) is restricted with $t \in [T_{IN-}, T_{IN+}]$. Projection outside the FOV is not acquired and for mathematical convenience, set to be 0 in $\mathbf{p}_{IN}(\theta,t)$.

$$\mathbf{p}_{IN}(\theta,t) = \begin{cases} \int_{-\infty}^{+\infty} \mathbf{\mu}_{FULL}(t\cos\theta - s\sin\theta, t\sin\theta + s\cos\theta) ds & t \in [T_{IN-}, T_{IN+}] \\ 0 & t \in (-\infty, T_{IN-}) \cup (T_{IN+}, +\infty) \end{cases} \quad (1)$$

Given the rays passing through $\boldsymbol{\mu}_{ROI}$ only, a complete ROI-only projection $\mathbf{p}_{ROI}$ is generated (see Eq. (2)) which can be well reconstructed by conventional algorithms. Unfortunately, in interior tomography, part of $\boldsymbol{\mu}_{EX}$ is also scanned which contributes an additional background projection $\mathbf{p}_{BG}$ (see Eq. (2)) to the projection inside the FOV. Hence, $\mathbf{p}_{IN}$ can be decomposed into two parts: a complete $\mathbf{p}_{ROI}$ from $\boldsymbol{\mu}_{ROI}$ and a truncated $\mathbf{p}_{BG}$ from $\boldsymbol{\mu}_{EX}$ (see Fig. 1(b)). If FBP is directly applied to $\mathbf{p}_{IN}$, reconstructions will suffer from obvious cupping artifacts and biased value due to truncated filtration.

$$\begin{aligned}
\mathbf{p}_{IN}(\theta,t) &= \mathbf{p}_{ROI}(\theta,t) + \mathbf{p}_{BG}(\theta,t) \\
\mathbf{p}_{ROI}(\theta,t) &= \int_{-\infty}^{+\infty} \boldsymbol{\mu}_{ROI}(t\cos\theta - s\sin\theta, t\sin\theta + s\cos\theta)ds \quad t \in [T_{IN-}, T_{IN+}] \\
\mathbf{p}_{BG}(\theta,t) &= \int_{-\infty}^{+\infty} \boldsymbol{\mu}_{EX}(t\cos\theta - s\sin\theta, t\sin\theta + s\cos\theta)ds \quad t \in [T_{IN-}, T_{IN+}]
\end{aligned} \quad (2)$$

Intuitively, $\boldsymbol{\mu}_{ROI}$ can be reconstructed from a non-truncated projection $\mathbf{p}_{FULL}$ (see the red arrow in Fig. 1(b)) with $\mathbf{p}_{IN}$ unchanged inside the original FOV $t \in [T_{IN-}, T_{IN+}]$ and an additional measure $\mathbf{p}_{EX}$ of $\boldsymbol{\mu}_{EX}$ outside the FOV $t \in (-\infty, T_{IN-}) \cup (T_{IN+}, +\infty)$:

$$\begin{aligned}
\mathbf{p}_{FULL}(\theta,t) &= \mathbf{p}_{IN}(\theta,t) + \mathbf{p}_{EX}(\theta,t) \\
\mathbf{p}_{EX}(\theta,t) &= \begin{cases} 0 & t \in [T_{IN-}, T_{IN+}] \\ \int_{-\infty}^{+\infty} \boldsymbol{\mu}_{EX}(t\cos\theta - s\sin\theta, t\sin\theta + s\cos\theta)ds & t \in (-\infty, T_{IN-}) \cup (T_{IN+}, +\infty) \end{cases}
\end{aligned} \quad (3)$$

Multiple methods have been proposed to acquire $\mathbf{p}_{EX}$ via extrapolation either heuristically or in a data-driven manner. The FOV extension is intended for the complete measure of $\boldsymbol{\mu}_{EX}$. However, compared with $\mathbf{p}_{EX}$, the measured information about $\boldsymbol{\mu}_{EX}$ in $\mathbf{p}_{BG}$ can be rather limited especially when the ROI is relatively small. This may make it difficult for accurate extrapolation and compromise the image quality.

As aforementioned, $\boldsymbol{\mu}_{ROI}$ can also be reconstructed from $\mathbf{p}_{ROI}$ (see the red arrow in Fig. 1(b)) which is mixed with $\mathbf{p}_{BG}$ in the truncated $\mathbf{p}_{IN}$. Here, we introduce a new pathway to reconstruct the ROI through extraction. Different from extrapolation, extraction only works inside the FOV to remove the contribution of $\boldsymbol{\mu}_{EX}$ and estimates the ROI-only projection $\hat{\mathbf{p}}_{ROI}$ without FOV extension. The proposed approach works in dual domains: a sinogram-domain network (SDNet) $\varphi_S(\cdot)$ estimates $\mathbf{p}_{BG}$ via extraction, an FBP layer for domain transformation, and an image-domain network (IDNet) $\varphi_I(\cdot)$ further improves the image quality. In practice, both SDNet and IDNet employ a modified U-shape network as the backbone and trained in a residual manner. The objective function can be formulated as

$$\hat{\vartheta}_{\varphi_S}, \hat{\vartheta}_{\varphi_I} = \arg\min_{\vartheta_{\varphi_S}, \vartheta_{\varphi_I}} \{\lambda \frac{\|\varphi_S(\mathbf{p}_{IN}) - \mathbf{p}_{ROI}\|_2}{\|\mathbf{p}_{ROI}\|_2} + \frac{\|\varphi_I(R^{-1}(\varphi_S(\mathbf{p}_{IN}))) - \boldsymbol{\mu}_{ROI}\|_2}{\|\boldsymbol{\mu}_{ROI}\|_2}\} \quad (4)$$

where $R^{-1}$ denotes the process of FBP reconstruction, $\vartheta_{\varphi_S}$ and $\vartheta_{\varphi_I}$ the network parameters for SDNet and IDNet respectively. $\lambda$ is simply set to 1 in the subsequent experiments.

## 3. EXPERIMENTS AND RESULTS

Interior tomography scanning is simulated under an equidistant fan-beam configuration. The full image is 512 × 512 with pixel size 0.5 mm where a 256 × 256 round region in the center is defined as the ROI. The detector bin size is 0.45 mm where 1056 elements were used for a fully non-truncated projection and 512 elements for the interior projection. The projection is acquired under 480 views uniformly distributed over $2\pi$. Poisson noise is simulated with $10^5$ incident photons per ray in the simulation. The simulation dataset is based on 39 patients from AAPM Low Dose Grand

Challenge and TCIA Low Dose CT Image and Projection Data are used with a split of 29-5-5 for training, validation, and test, respectively. Structural similarity index (SSIM) and peak signal to noise ratio (PSNR) in the test set are used for quantitative evaluation.

We employed four methods as baselines for evaluation: the conventional FBP reconstruction, FBPConvNet, a dual-domain network where the sinogram-domain only works on extrapolation (DD_Ex), and a dual-domain network in[12] where the sinogram-domain network employs two heads output for extrapolation outside the FOV and denoising inside the FOV (DD_ExDenoise) respectively.

Two samples from different patients are displayed in Fig. 2 for demonstration. FBP reconstructions are contaminated by severe attenuation value bias and noise. FBPConvNet can recover the gray level but the noise cannot be well suppressed. Compared with FBPConvNet, DD_Ex achieves better noise reduction but the images are blurry which compromises context structures (see the yellow arrows in Fig. 2). DD_ExDenoise can suppress the cupping artifacts and noise to a large extent. But some inconsistent artifacts can be observed (see the blue circle in Fig. 2). Above all, the proposed extraction-based methods can produce the most reliable structures, especially in the low-contrast regions (see the zoom-ins in Fig. 2). The quantitative metrics among all the test slices are listed in Table I. We can find that the proposed method achieves the best SSIM while the mean PSNR is slightly inferior to DD_ExDenoise.

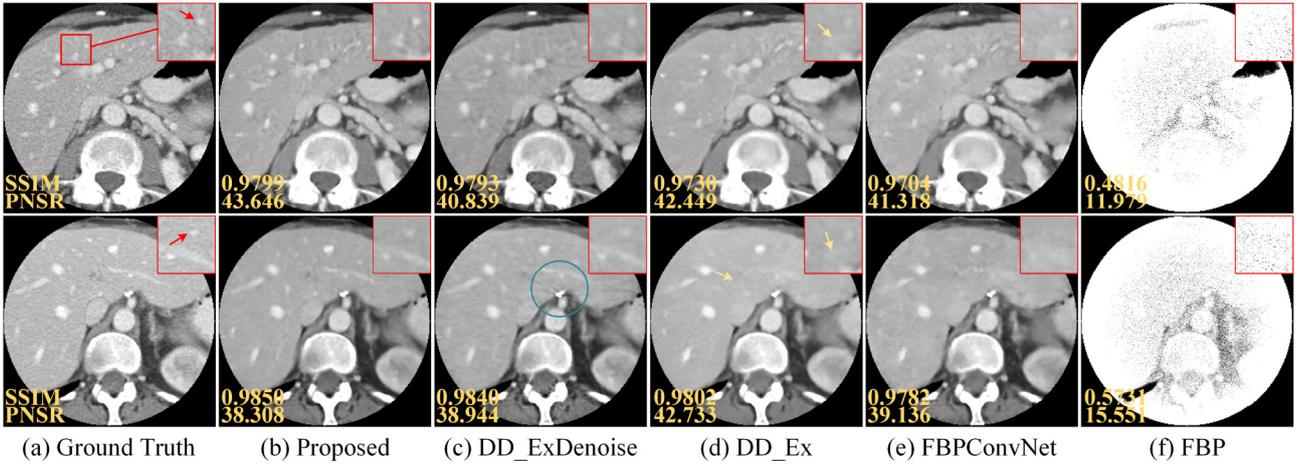

(a) Ground Truth    (b) Proposed    (c) DD_ExDenoise    (d) DD_Ex    (e) FBPConvNet    (f) FBP

**Fig. 2** Comparison among different methods: ground truth (a), proposed method (b), DD_ExDenoise (c), DD_Ex (d), FBPConvNet (e), and FBP (f). The display window is [0.016, 0.024]. SSIM and PSNR are in the bottom-left corner of each image.

Table I
Quantitative Metrics among all Test Slices

| Incident photons | $10^5$ | | $5 \times 10^4$ | |
|---|---|---|---|---|
| Metric | SSIM | PSNR | SSIM | PSNR |
| FBP | 0.5363 | 12.402 | 0.4594 | 12.317 |
| FBPConvNet | 0.9509 | 38.475 | 0.9327 | 32.855 |
| DD_Ex | 0.9551 | 39.881 | 0.9455 | 38.291 |
| DD_ExDenoise | 0.9644 | **40.327** | 0.9513 | 38.830 |
| Proposed | **0.9676** | 40.150 | **0.9549** | **38.964** |

To further evaluate the generalization performance, without finetuning, we directly evaluated the models on the same test set phantom but injected higher noise (corresponding to $5 \times 10^4$ photons) in projection. The same two slices are shown in Fig. 3 and the quantitative metrics are listed on the right side of Table I. FBPConvNet cannot adapt to the higher noise and produces images with obvious gray level deviation. Though DD_Ex may achieve better noise reduction and achieve higher PSNR, some low-contrast structures are incorrect or distorted (see the yellow arrows in Fig. 3). It can be observed that DD_ExDenoise may produce some streaking and inconsistent artifacts due to the discrepancy in the noise level. The proposed method achieves the best image quality with fewer artifacts and fake structures than other methods, which validated its capability of generalizing to higher noise levels.

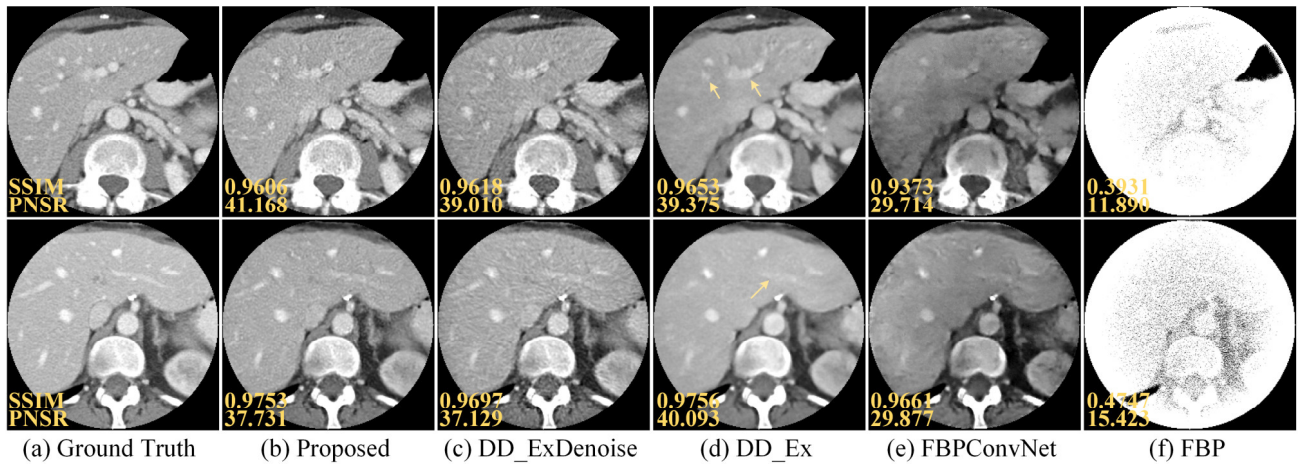

(a) Ground Truth (b) Proposed (c) DD_ExDenoise (d) DD_Ex (e) FBPConvNet (f) FBP

**Fig. 3** Comparison among different methods on dataset injected higher noise: ground truth (a), proposed method (b), DD_ExDenoise (c), DD_Ex (d), FBPConvNet (e), and FBP (f). The display window is [0.016, 0.024]. SSIM and PSNR are in the bottom-left corner of each image.

### 4. DISCUSSION AND CONCLUSION

In this paper, we established a new pipeline for interior tomography that works on extraction. The main difference between extraction- and extrapolation- based pathways is how to alleviate the influence of the exterior region which is not fully measured. In the literature, extrapolation-based methods have been proposed to extend the truncated interior projection to acquire a non-truncated projection that contains the full measure of the exterior region. However, extraction learns to estimate the contribution of the exterior region inside the FOV to achieve an ROI-only projection without sinogram extension. This pipeline requires less prior information about the exterior region which may make the reconstruction more stable. Experiments have favored the extraction-based method in its potential of revealing reliable structures and good generalization. Besides, as the extraction-based method only works inside the FOV, it can be more capable of jointly dealing with other nonideal factors such as noise in the interior projection compared with extrapolation-based methods (see the comparison with DD_ExDenoise and DD_Ex in Fig. 2 and Fig. 3). Although the proposed method has shown promising results for interior reconstruction, slight structural distortions and gray level error may still exist. We will conduct a more detailed analysis in our future work.